\documentstyle[11pt,mrs2001,epsfig]{article}
\begin{document}
\title{PROBING THE MASS DISTRIBUTION WITH IRAS GALAXIES}

\author{E. Branchini $^1$}
\affil{$^1$  Dipartimento di Fisica, Universit\'a di Roma TRE,
Via della Vasca Navale 84, 00146, Roma, Italy}

\begin{abstract}
We present the results of three 
independent analyses in which we show that IRAS galaxies 
faithfully trace the underlying distribution the mass
in the local universe.
In the first analysis we 
compare the mass and the galaxy density fields
and show that they are consistent within 60 $h^{-1}$Mpc.
In the second one  we show that the tidal velocity field of 
the Mark III catalog is consistent with the tidal velocity  
field predicted from the distribution of IRAS galaxies, 
hence indicating that IRAS galaxies trace 
the mass distribution well beyond 60 $h^{-1}$Mpc.
Finally, a third analysis aimed at
determining the mean biasing relation of 
IRAS galaxies showed no
appreciable deviations from the linear biasing prescription.

\end{abstract}

\section{Introduction}

IRAS redshift surveys are characterized by 
their considerable depth and almost complete sky coverage and thus 
have been extensively used to investigate  the mass distribution in the 
local universe.
As a result, most of the local cosmological information 
is derived under the assumption that IRAS galaxies are faithful tracers
of the underlying mass distribution. Therefore, it is of fundamental 
importance to test the validity this hypothesis and to
quantify the so called biasing relation, i.e. the 
map between the mass  and the galaxy spatial distributions
both smoothed with the same filter and on the
same smoothing scale.

This problem has been already addressed in several different ways  by
a number authors.
Baker {\it et al.} (1998) and Seaborne {\it et al.} (1999) have shown
that distribution of IRAS galaxies 
is consistent with that of optical galaxies on scales larger that
5 $h^{-1}$Mpc. This indicates that  different types of objects trace a
common underlying density field.
More important, direct comparisons with the mass density field
obtained from the observed peculiar velocities of galaxies using POTENT-like 
algorithms (Bertschinger \& Dekel 1989) or Wiener Filtering techniques
(Zaroubi  {\it et al.} 1995), have shown that the distribution of IRAS 
galaxies is consistent with that of the mass  within a region of  
$\approx 60 h^{-1}$Mpc (see Sigad {\it et al.} 1998 and reference therein).

Our aim is to improve the accuracy of the previous
analyses, extend them to scales larger than  60 $h^{-1}$Mpc
and determine whether the biasing of IRAS galaxies is consistent
with a linear relation.

\section{IRAS Galaxies vs. Mass within 60 $h^{-1}$Mpc}

\begin{figure}
\vspace{10truecm}
\centering
{\includegraphics{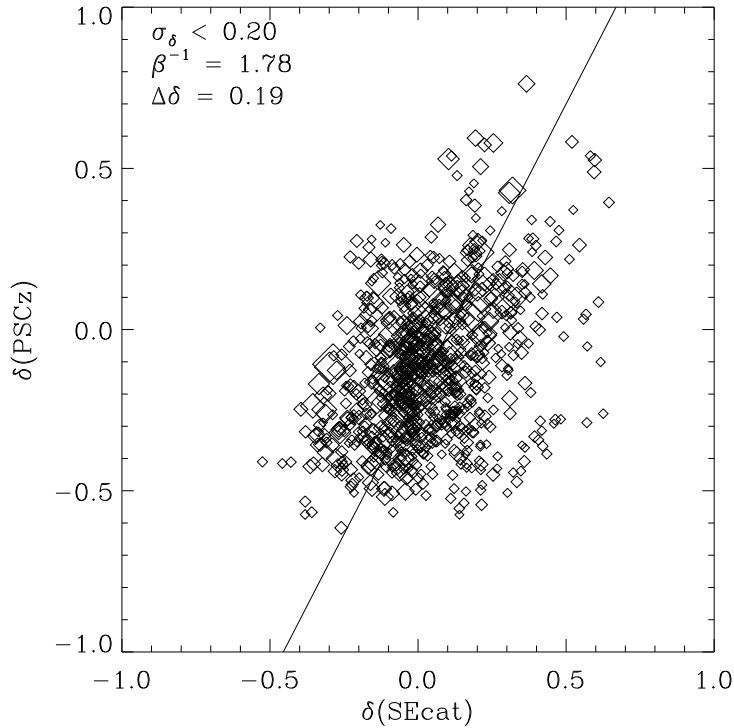}}
\caption{
Scatterplot showing the UMV reconstructed mass density {it vs.}
the PSC$z$ galaxy density. The size of the symbols is inversely
proportional to the errors}
\label{fig:1}
\end{figure}

In the first analysis we perform a density-density comparison
i.e. we reconstruct the mass density field 
within 60 $h^{-1}$Mpc from observed galaxy peculiar velocities
and compare it with the 3D galaxy distribution obtained
from the IRAS redshift surveys.
In our analysis we use a new reconstruction method, called
Unbiased Minimal Variance estimator [UMV, hereafter],
a new catalog of peculiar velocity, called SEcat, and
the most recent redshift survey of IRAS galaxies, called PSC$z$.

The UMV estimator proposed by Zaroubi (2001) 
can be viewed  as a compromise between POTENT algorithm and the Wiener filtering
technique. It constitutes an improvement over both these methods
since it provides an unbiased minimum variance reconstruction of the underlying density field 
from incomplete, noisy and sparse catalogs of galaxy peculiar
velocities and also account for the correlation properties of the 
underlying velocity field.  
The recent SEcat catalog of peculiar velocities results from the merging of 
the SFI catalog of spiral galaxies (Giovanelli  {\it et al.} 1997)
and the ENEAR catalog of early type galaxies (da Costa  {\it et al.} 2000). 
The large number of objects,
homogeneous sky coverage and composite nature of the catalog guarantee 
dense, uniform and unbiased sampling of the cosmic velocity field.
Finally, the PSC$z$ catalog  (Saunders 1996) 
constitutes the largest and deepest all-sky redshift survey of IRAS 
galaxies to date.

The preliminary results of the
comparison between the UMV-reconstructed mass 
and the IRAS PSC$z$ galaxy density fields are shown in Fig. 1.
Each point displayed in the scatter-plot shows the comparison 
between the UMV and PSC$z$ overdensities smoothed with a Gaussian filter 
of radius 12  $h^{-1}$Mpc and measured at the same location.
The correlation between the two fields  
is linear and thus consistent with 
a biasing relation on the scale of the smoothing.
The slope of the solid line returns $\beta \equiv \Omega^{0.6}/b=0.56^{+0.12}_{-0.10}$, where
$\Omega$ and $b$ indicate the mass density and linear bias parameters, respectively. 
The distribution of the density residuals is consistent with Gaussian, 
hence indicating that the PSC$z$ density field is an adequate model 
for the underlying mass density field within a radius of
60 $h^{-1}$Mpc  under the assumption of linear biasing
and on a scales larger than the smoothing length. 
It is worth stressing that, due to the regularization properties of 
the UMV estimator and the unprecedented quality of the 
SEcat and PSC$z$ datasets, the statistical significance of 
this density-density match is greater than 
in previous comparisons of this type.

The UMV estimator can be also used to reconstruct the underlying continuous velocity 
field. The comparison with the linear model velocity field predicted from the distribution of 
PSC$z$ galaxies  returns a value of 
$\beta=0.51^{+0.04}_{-0.04}$, consistent with the one obtained from densities
(Zaroubi {\it et al. in prep.}).

\begin{figure}
\vspace{8truecm}
%\centering
{\includegraphics{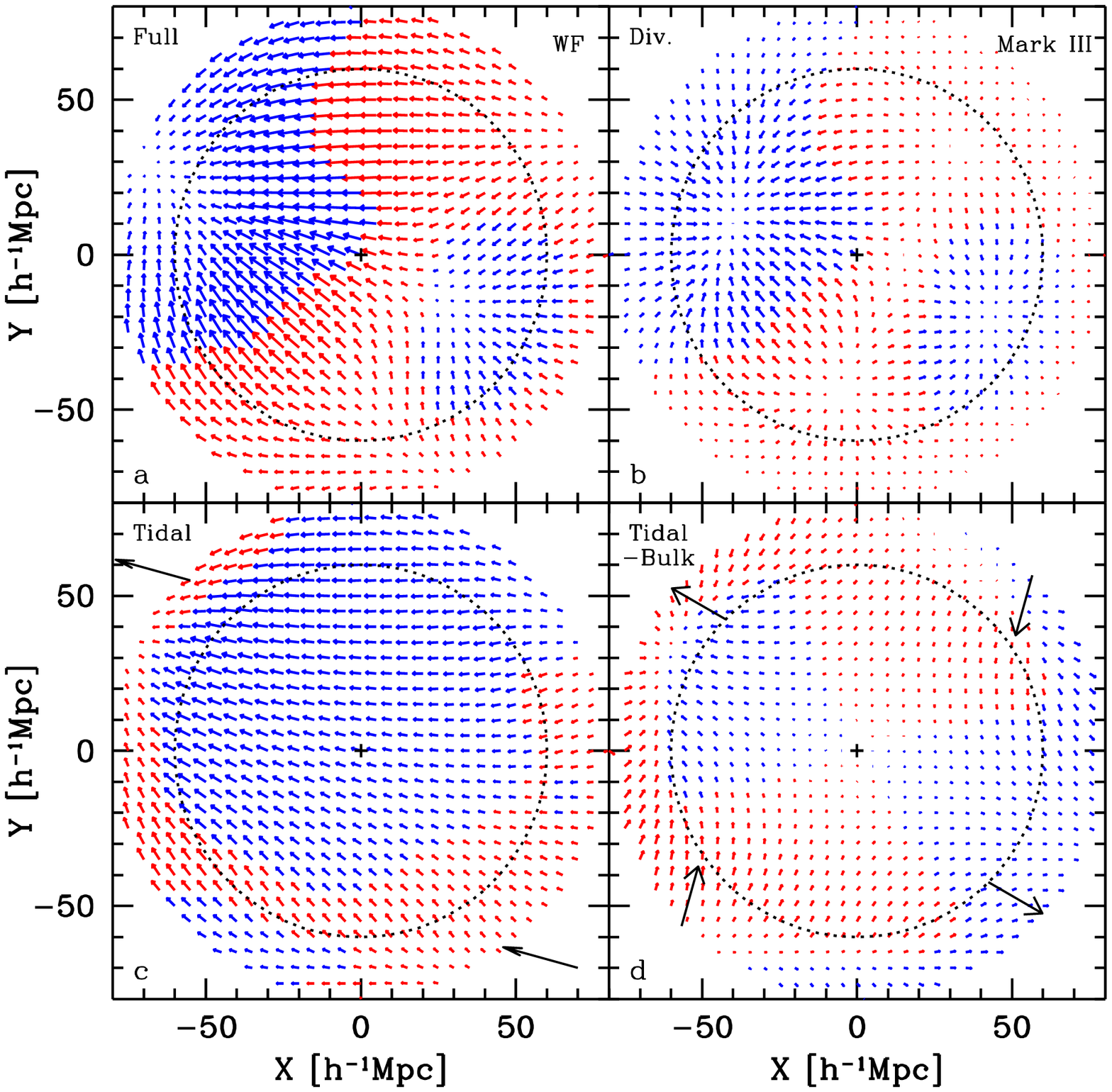}}
\caption{Decomposition of the MARK III velocity field in the 
Supergalactic plane in a sphere of  60 $h^{-1}$Mpc about the Local Group.
Top left: full velocity field. Top Right: divergent component.
Bottom left: Tidal component. Bottom right: Tidal component - dipole moment.
The long arrows show the directions of the dipole 
(bottom-left) and of the
dilational and compressional eigenvectors of the shear
(bottom-right).}
\vspace{11.truecm}
{\includegraphics{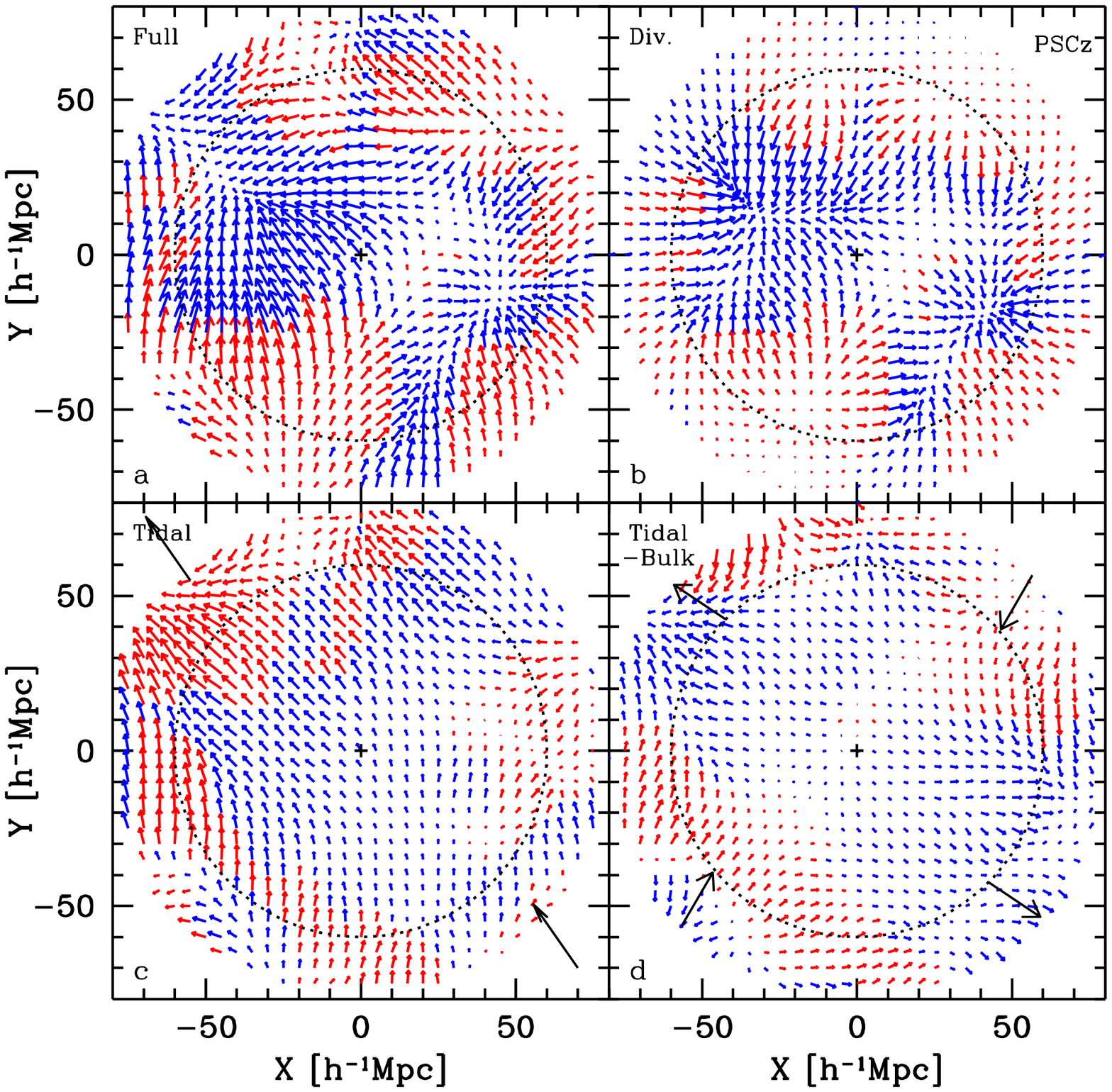}}
%\vspace{.5truecm}
\caption{Same as Fig. 2 but for the PSC$z$ linear velocity field.}
%\label{fig:2}
%\end{figure}
%\begin{figure}
%\label{fig:3}
\end{figure}

\section{IRAS Galaxies vs. Mass beyond 60 $h^{-1}$Mpc}

To determine the mass distribution on scales larger than 
60 $h^{-1}$Mpc and compare it to that of IRAS galaxies we have 
used the tidal field decomposition technique proposed 
by Hoffman {\it et al.} (2001).

In this method the Wiener filtering technique
is applied to the Mark III peculiar velocity catalog
(Willick {\it et al.} 1997)
to reconstruct the full density and velocity fields  
(top-left panel of Fig. 2). The density field 
within 60 $h^{-1}$Mpc is used to determine, via linear theory,
the divergent part of the velocity field (top-right).
The tidal velocity field (bottom-left) is obtained by subtracting 
the divergent component to the full velocity field. This tidal field 
is only determined by the mass distribution beyond 60 $h^{-1}$Mpc
and is dominated by its dipole and quadrupole moments.
The velocity field shown in the bottom-right panel was obtained after 
subtracting the dipole component from the tidal field and is 
clearly characterized by a quadrupole pattern. 
The direction of the Mark III dipole vector
is displayed as long arrows in 
the bottom-left panel of Fig. 2.
The long arrows in the bottom-right panel
show the directions of the 
compressional and dilational eigenvectors of the shear tensor.

A similar decomposition was applied to the linear velocity field 
obtained from the distribution of IRAS PSC$z$ galaxies.
The full velocity field and its divergent part are obtained, via linear theory,
from the  distribution of PSC$z$ galaxies within 200  $h^{-1}$Mpc
and 60 $h^{-1}$Mpc, respectively (upper plots of Fig. 3).
The PSC$z$ tidal velocity field, obtained from the galaxy distribution 
between 60 and 200  $h^{-1}$Mpc is shown
in the two bottom panels with its dipole component (bottom-left) 
and without it (bottom-right).
As in Fig. 2, the long arrows shown in Fig. 3
indicate the directions of dipole (bottom-left)
and shear eigenvectors (bottom-right) 
of the PSC$z$ tidal velocity field.

The preliminary results of our quantitative analysis indicate that the 
dipole and quadrupole moments of the Mark III and PSC$z$ tidal velocity fields 
agree within the errors (Hoffman {\it et al.} {\it in prep.}).
This match strongly suggests that the distribution
of the mass beyond 60 $h^{-1}$Mpc is consistent with that traced by 
IRAS galaxies, at least in term of dipole and quadrupole moments.

\section{The Mean Biasing Function of IRAS galaxies}

Finally, we have used the method of Sigad, Branchini \& Dekel (2000)
to determine the mean biasing function of IRAS galaxies from their 
observed one point probability distribution function [PDF]. The method assumes that 
the bias is deterministic and monotonic and that the the mass PDF is 
lognormal on the scales of interest.
To apply this technique we have extracted a
volume limited subsample from the PSC$z$ galaxy survey
and applied a spherical Top Hat filter of
radius  8 $h^{-1}$Mpc to estimate the PDF of galaxies.
Then the mean biasing function has been determined by 
mapping the galaxy PDF into that of the mass, modeled by 
a lognormal distribution.
Extensive numerical tests have shown that the method is very 
effective to measure deviations from linear biasing.
However, the slope of the mean biasing function is sensitive
to the rms density fluctuations, $\sigma_8$, which 
fully characterizes a lognormal distribution and needs 
to be assumed {\it a priori}.

\begin{figure}
\vspace{10truecm}
%\centering
{\includegraphics{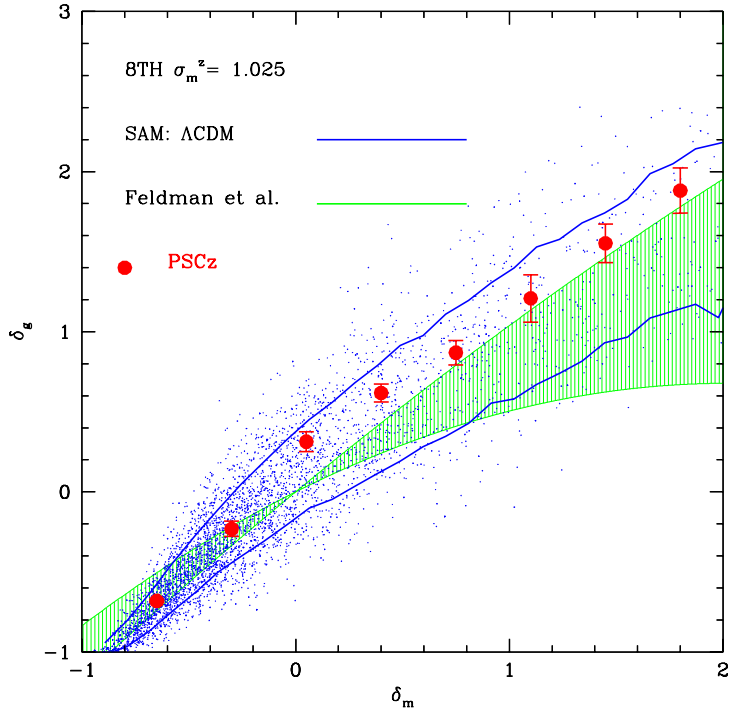}}
\caption{Large dots: PSC$z$ mean biasing function. The results refer to a
Top Hat smoothing of 8 $h^{-1}$Mpc and to $\sigma_8=1.025$
The two solid lines show the expected stochasticity
of the bias relation from the SAM predictions.
The shaded region shows the biasing relation 
determined by Feldman (2000) }
\label{fig:4}
\end{figure}

The preliminary results are shown in Fig. 4.
The large dots show the PSC$z$ mean biasing function 
obtained assuming $\sigma_8=1.025$. The function is clearly 
consistent with a linear biasing scheme.
The two solid lines indicate the $1 \sigma$ uncertainty interval
around the mean biasing,
associated to the stochasticity of the biasing relation.
as predicted by the Semianalytic Model [SAM] of Somerville {\it et al.} (2000).
The dashed region is plotted for reference and 
shows the  biasing function of IRAS galaxies determined
by Feldman  {\it et al.} (2001) that, 
however, refers to scales 
much larger than those of our analyses.

\section{Conclusions}

The preliminary results presented here corroborate the 
common assumption that IRAS galaxies are faithful tracers 
of the mass distribution.
In the first analysis we have applied the new UMV estimator to show that 
the distribution of IRAS galaxies matches that of the mass 
within  60 $h^{-1}$Mpc. This  confirms the results of similar
comparisons previously performed using a number of different techniques and datasets.
In the second analysis we take advantage of the nonlocal properties
of the galaxy peculiar velocities to infer the mass distribution
beyond  60 $h^{-1}$Mpc and found that is consistent with that
of IRAS PSC$z$ galaxies.
Finally, we use the inverse-mapping method of Sigad, Branchini \& Dekel (2000)
to quantify the biasing relation between galaxies and mass.
Combined with the result of the likelihood analysis of Tegmark {\it et al.} 
(2000), our results show that on a scale of 
8 $h^{-1}$Mpc  IRAS PSC$z$ galaxies obey a mean linear biasing relation
with a biasing parameter $b$ consistent with unity.

\acknowledgements{EB is grateful to Y. Hoffman and S. Zaroubi
for kindly providing  unpublished results and figures.}

\vfill
\end{document}